\documentclass{article}

\usepackage{PRIMEarxiv}

\usepackage[utf8]{inputenc}
\usepackage[T1]{fontenc}
\usepackage{hyperref}
\usepackage{url}
\usepackage{booktabs}
\usepackage{amsmath,amssymb,amsfonts}
\usepackage{nicefrac}
\usepackage{microtype}
\usepackage{xcolor}
\usepackage{textcomp}
\usepackage{float}
\usepackage{fancyhdr}
\usepackage{graphicx}
\usepackage[numbers,sort&compress]{natbib}
\graphicspath{{./}}
\hypersetup{colorlinks=true,linkcolor=blue,citecolor=blue,urlcolor=blue}

\title{CalibCause: A Quality-Gated, Interpretable and Calibrated Deep Learning Framework for Diabetic Retinopathy Grading}

\author{
  Sohaib Sulman\thanks{Corresponding author.} \\
  Department of Data Science and \\
  Artificial Intelligence \\
  National University of Computer and \\
  Emerging Sciences (FAST-NUCES) \\
  Islamabad 44000, Pakistan \\
  \texttt{i248068@isb.nu.edu.pk} \\
  \And
  Muhammad Atif Saeed \\
  Department of Data Science and \\
  Artificial Intelligence \\
  National University of Computer and \\
  Emerging Sciences (FAST-NUCES) \\
  Islamabad 44000, Pakistan \\
  \texttt{atif.saeed@isb.nu.edu.pk} \\
  \AND
  Akhtar Jamil \\
  Department of Computer Science \\
  National University of Computer and \\
  Emerging Sciences (FAST-NUCES) \\
  Islamabad 44000, Pakistan \\
  \texttt{akhtar.jamil@nu.edu.pk} \\
  \And
  Alaa Ali Hameed \\
  Department of Computer Engineering \\
  Istinye University \\
  Istanbul 34408, Turkey \\
  \texttt{alaa.hameed@istinye.edu.tr} \\
  \AND
  Roland Aydin \\
  German Research Center for \\
  Artificial Intelligence (DFKI) \\
  Saarbr\"ucken, Germany \\
  \texttt{roland.aydin@uni-saarland.de} \\
}

\begin{document}
\maketitle

\begin{abstract}
\textbf{PURPOSE:} Most deep learning systems for diabetic retinopathy (DR) grading are reported on one data source and say little about how they behave on a new camera or grading team, how sure they are of a given grade, or why they chose it. We built CalibCause, a quality-gated mixture-of-experts grader that adds per-expert visual explanations, a quality signal that needs no manual labels, a compact distilled student, and calibrated confidence with deferral. We tested it against a plain ConvNeXt-Tiny trained with the same recipe.

\textbf{METHODS:} We used 51,310 fundus images from EyePACS, DDR and APTOS~2019, treated as three acquisition environments, and IDRiD and DeepDRiD for external testing. Models were compared on a pooled test split and by leave-one-environment-out (LOEO) testing, with paired bootstrap intervals, a second training seed, and a diagnostic that refits the four grade thresholds on 200 labelled images from the new site.

\textbf{RESULTS:} On the pooled test split the label-free student (27.8M parameters, 52\% fewer than its teacher) reached quadratic weighted kappa (QWK) 0.863 and accuracy 0.865. In LOEO testing the mean QWK of the label-free model (0.741) was close to that of the plain baseline (0.743); the two were not distinguishable on DDR and APTOS, and the baseline was higher on held-out EyePACS (0.547 vs.\ 0.533). The adversarial invariance block did not improve held-out accuracy. Refitting the grade thresholds on 200 images raised held-out EyePACS QWK from 0.533 to 0.621 for the label-free model and from 0.524 to 0.597 for the full model, but only from 0.547 to 0.565 for the plain baseline. After vector scaling, expected calibration error was 0.025--0.036 for all five CalibCause models, and deferring the least confident 20\% of images raised the student's accuracy from 0.865 to 0.923.

\textbf{CONCLUSION:} CalibCause matches a plain ConvNeXt-Tiny in cross-site accuracy while adding explanations, label-free quality conditioning, a compact student and a usable confidence signal. It does not make grading site-invariant by itself. Part of the loss on a new site is a shift in the grade thresholds, which a small local sample partly repairs.
\end{abstract}

\keywords{Diabetic retinopathy \and Mixture of experts \and Domain generalisation \and Knowledge distillation \and Calibration \and Explainability}

\section{Introduction}\label{sec:intro}

Diabetic retinopathy (DR) remains one of the main causes of preventable vision loss in working-age adults, and regular fundus photography is how most screening programmes find it early. Deep learning has made automated grading of these photographs practical~\cite{gulshan2016,ting2017,grzybowski2020ai,li2022eyedr}, and models trained on curated images now come close to human graders. Photographs from routine screening are messier, with poor illumination and blur, and a model trained mostly on clean images may behave unpredictably on them.

Image quality is also tied to disease. QGDR~\cite{pan2026qgdr} describes this as quality--severity coupling: poorer images tend to belong to patients with more advanced disease. In their data the share of referable DR rose from 17.4\% to 27.8\% as image quality fell on EyeQ~\cite{fu2019eyeq}, and from 28.6\% to 81.9\% on DDR. A quality filter therefore discards the images most likely to come from patients who need referral. QGDR instead uses quality as a signal that conditions the grader, in line with earlier quality-aware work~\cite{hou2023cleaqdr,che2023mkcnet}.

That design leaves several practical questions open. It was trained within one data source at a time and tested on two external sets, but no training source was held out in turn, so it says little about how a model trained on several sites transfers to a new camera, clinic or grading team. Site-to-site variation is a known cause of failure in DR grading~\cite{zhu2025irm,bi2025gad} and in medical imaging more broadly~\cite{zech2018}. Its multi-expert structure is harder to inspect than a single network, and a compact version was left to future work, although distillation has been used for DR classification~\cite{dejene2026distillation}. Its quality signal came from human quality labels, which are costly to collect: even recent continuous quality scores rely on ratings by ophthalmologists~\cite{gong2025fthnet}. Finally, DR grades depend on who assigned them: grader variability and the reference standard change what a model appears to learn~\cite{krause2018}, even under the same international severity scale~\cite{wilkinson2003}.

We built CalibCause to address these questions in one fundus-only framework. It adds an environment-adversarial block that discourages the shared representation from encoding the data source~\cite{ganin2016dann}; per-expert visual explanations~\cite{gradcam}; a label-free quality gate; a distilled single-stream student~\cite{hinton2015kd}; and post-hoc calibration~\cite{guo2017calibration} used to defer uncertain cases. We evaluated all of these against a plain ConvNeXt-Tiny baseline trained with the same recipe, on a pooled split, with each source held out in turn, and on two external datasets.

The comparison with the plain baseline shapes what we can and cannot claim. CalibCause matched it on cross-site accuracy but did not beat it, and the adversarial block did not help held-out accuracy. What it adds is what a plain classifier does not provide by itself: inspectable routing and Grad-CAM outputs, a quality signal that needs no manual labels, a student half the size of the teacher that outperformed it, and a confidence score good enough to support deferral. We also show that part of the large drop on a held-out site is a shift in where the grade boundaries fall, which can be repaired with a small local sample.

\section{Subjects and Methods}\label{sec:methods}

\subsection*{Datasets}
We used 51,310 fundus images from three public screening cohorts, each treated as a separate acquisition environment: EyePACS~\cite{eyepacs2015} (35,126 images from a multi-clinic screening programme), DDR~\cite{li2019ddr} (12,522 images from 147 hospitals in China) and APTOS~2019~\cite{aptos2019} (3,662 images from an Indian screening programme). All images carry the five-point DR grade (0--4) and were resized to $384\times384$ pixels. Splits were stratified jointly by environment and grade, giving 35,916, 5,131 and 10,263 images for training, validation and testing (Table~\ref{tab:data}). IDRiD~\cite{idrid} (516 images) and DeepDRiD~\cite{deepdrid} (400 images) served as external test-only sets: no image or label from either was used for training or tuning. In total the study used 52,226 images from five public datasets.

\begin{table}[!tbp]
\caption{Data used in the study. Test counts for the three development environments refer to the pooled test split.}\label{tab:data}
\small
\begin{tabular}{@{}lrl@{}}
\toprule
Dataset & Images & Role \\
\midrule
EyePACS & 35,126 (test: 7,026) & Development environment \\
DDR & 12,522 (test: 2,504) & Development environment \\
APTOS 2019 & 3,662 (test: 733) & Development environment \\
IDRiD & 516 & External test only \\
DeepDRiD & 400 & External test only \\
\bottomrule
\end{tabular}
\end{table}

\subsection*{Model}
\begin{figure}[!tbp]
\centering
\includegraphics[width=\linewidth]{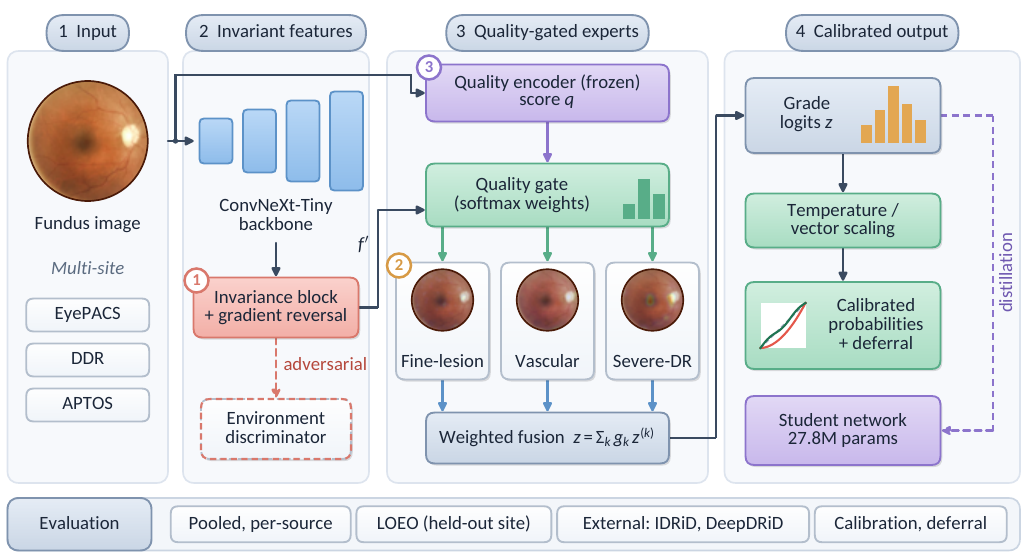}
\caption{Overview of CalibCause. A fundus image passes through a shared backbone and an environment-invariant transform (Module~1, with an adversarial environment discriminator). A separate frozen encoder scores the image quality, and this score gates three experts whose outputs are fused into logits that are then calibrated. Module~2 provides per-expert Grad-CAM maps and a distilled single-stream student. Module~3 is the label-free quality score. The fundus image and the per-expert Grad-CAM thumbnails are a real test image processed by the label-free model (more examples in Fig.~\ref{fig:cam}); the bar and curve icons are schematic.}\label{fig:arch}
\end{figure}

Figure~\ref{fig:arch} outlines the framework. CalibCause builds on the quality-gated mixture-of-experts design of QGDR~\cite{pan2026qgdr}, in line with earlier quality-aware diagnosis~\cite{che2023mkcnet}. A ConvNeXt-Tiny backbone~\cite{convnext2022} extracts features, which pass through an environment-invariant transform (Module~1). The transformed features are gated by an image-quality score and routed to three experts specialised for fine lesions, vascular lesions and severe disease; a weighted fusion of the experts gives the final grade. The grading loss is label-smoothed, class-weighted cross-entropy plus an ordinal penalty that grows with the squared distance between predicted and reference grade, which matches how QWK scores errors.

Module~1 trains an environment discriminator on the invariant features through a gradient-reversal layer~\cite{ganin2016dann}, so the shared representation is pushed away from encoding which source an image came from. Module~2 computes Grad-CAM~\cite{gradcam} for each expert, scales each map by that expert's gate weight for the image, and distils the trained teacher into a single-stream student~\cite{hinton2015kd} that drops the quality branch, gate and experts. Module~3 replaces the calibrated quality score with the raw output of a self-supervised quality encoder, so no human quality label is used at any point. We call this the label-free variant. Three further models were trained with the same recipe: the full model (calibrated quality score, Modules~1 and~2), a version of the full model without the invariance block and environment discriminator, and a plain ConvNeXt-Tiny with the same ordinal-aware loss, optimiser and epoch budget but no quality branch, gate, experts or distillation. The student has the same architecture as the plain baseline (27.82M parameters); the teacher has 58.21M.

\subsection*{Training}
The quality encoder was pretrained self-supervised for 15 epochs; the full model additionally calibrated its quality head on EyeQ labels for 10 epochs, and the label-free model did not. Both were then frozen. The grading networks were trained for 30 epochs with AdamW~\cite{adamw2019} (learning rate $3\times10^{-4}$ for new layers and $3\times10^{-5}$ for the pretrained backbone, weight decay 0.05), mixed precision, gradient clipping at 5.0 and a batch size of 6, using the \texttt{timm} library~\cite{timm}. The environment discriminator was weighted at 0.3 and phased in after the first epoch. A small entropy penalty (target 0.6) discouraged the gate from collapsing onto one expert. The plain baseline used the same settings. All development models used seed 42; a second seed (43) was run for the plain, full and label-free models on the held-out EyePACS fold.

The students were trained with a distillation loss of 0.7 times the Kullback--Leibler divergence to the teacher's softened outputs (temperature 6) plus 0.3 times the ordinal-aware cross-entropy. The label-free student was trained for 15 epochs with learning rates of $1.5\times10^{-4}$ (new head) and $1.5\times10^{-5}$ (pretrained backbone), a 300-iteration warm-up and gradient clipping at 5.0; a single rate of $3\times10^{-4}$ on all layers had collapsed to predicting grade~0 for every image. The student of the full model was trained earlier with a single rate of $3\times10^{-5}$.

\subsection*{Evaluation protocols}
In the \emph{pooled} protocol a model was trained and tested on the three-environment split. In the \emph{per-source} protocol it was trained and tested on one environment, following QGDR. In the \emph{leave-one-environment-out} (LOEO) protocol it was trained on two environments and tested on the third, which was never seen in training. Quadratic weighted kappa (QWK) was the primary metric. Accuracy and, for referable DR (grade $\ge2$), sensitivity, specificity and area under the ROC curve (AUC) were also computed at each model's default operating point.

\subsection*{Statistical analysis}
Ninety-five percent intervals for QWK were obtained by non-parametric bootstrap of the test images (2,000 resamples). Differences between models were assessed with a paired bootstrap in which the same resampled images were used for both models, and intervals on the difference that exclude zero were treated as evidence of a real difference on that test set. These intervals reflect test-set sampling only; they do not capture variation from retraining with a different seed, which we examined separately on the held-out EyePACS fold.
Secondary metrics were computed from the uncalibrated softmax outputs and all are reported: referable-DR AUC (score: summed probability of grades 2--4), sensitivity at 90\% specificity, macro-F1, mean absolute grade error (MAE), the fraction graded within one grade of the reference, and negative log-likelihood (NLL).

\subsection*{Threshold-refit diagnostic}
To ask how much of the drop on a new site is a shift in the grade boundaries rather than a loss of useful features, we replaced the argmax rule by the expected grade $\sum_k k\,p_k$ with four ordered cut-points. The cut-points were fitted by differential evolution~\cite{storn1997} to maximise QWK on a random sample of 200 labelled images drawn from the validation split of the held-out site, never from its test split, and the procedure was repeated for 30 random draws. As a control we ran the same procedure on the in-domain model and the EyePACS test images; the change there measures the effect of the decision rule alone.

\subsection*{Calibration and deferral}
Expected calibration error (ECE) used 15 equal-width confidence bins. Temperature scaling and vector scaling (a per-class scale and bias on the logits) were fitted on the validation split by minimising negative log-likelihood and applied unchanged to the test split. The same procedure was applied to every model, so ECE values can be compared directly. Deferral ranked test images by temperature-scaled confidence and reported accuracy on the most confident fraction (the coverage).

\section{Results}\label{sec:results}

\subsection*{Pooled grading and the distilled student}
Table~\ref{tab:pooled} gives results on the pooled test split. The label-free student was the most accurate of the five CalibCause models (accuracy 0.865, QWK 0.863, referable AUC 0.969) and exceeded its own teacher (QWK 0.839) with 52\% fewer parameters. The student of the full model behaved similarly (QWK 0.840 against 0.813 for its teacher). The model without the invariance block was the best of the teachers (QWK 0.861).

\begin{table}[!tbp]
\caption{Results on the pooled test split ($n=10{,}263$). QWK with 95\% bootstrap interval; AUC is for referable DR (grade $\ge2$). ECE is shown without calibration (raw), after temperature scaling (T) and after vector scaling (V), all fitted on the validation split. Parameters (Par.) in millions.}\label{tab:pooled}
\footnotesize
\setlength{\tabcolsep}{2.6pt}
\begin{tabular}{@{}lcccccccc@{}}
\toprule
Model & Par. & Acc & QWK (95\% CI) & AUC & ECE raw & ECE T & ECE V \\
\midrule
Full, teacher & 58.21 & 0.829 & 0.813 (0.800--0.825) & 0.938 & 0.270 & 0.144 & 0.031 \\
Label-free, teacher & 58.21 & 0.849 & 0.839 (0.827--0.850) & 0.945 & 0.261 & 0.149 & 0.025 \\
No invariance block, teacher & -- & 0.857 & 0.861 (0.851--0.870) & 0.944 & 0.223 & 0.117 & 0.030 \\
Full, student & 27.82 & 0.851 & 0.840 (0.829--0.851) & 0.968 & 0.182 & 0.057 & 0.036 \\
Label-free, student & 27.82 & 0.865 & 0.863 (0.853--0.873) & 0.969 & 0.173 & 0.055 & 0.033 \\
\bottomrule
\end{tabular}
\end{table}

\subsection*{Cross-site accuracy against a plain baseline}
\begin{figure}[!tbp]
\centering
\includegraphics[width=0.88\linewidth]{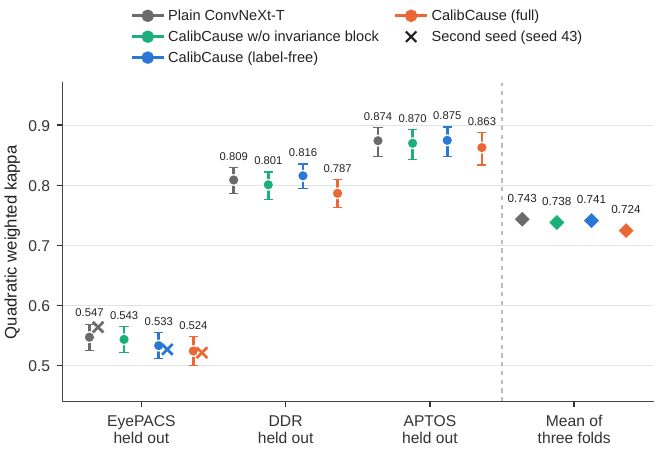}
\caption{Leave-one-environment-out QWK for the four models. Points show seed 42 with 95\% bootstrap intervals; crosses on the EyePACS fold show a second training seed.}\label{fig:loeo}
\end{figure}

\begin{figure}[!tbp]
\centering
\includegraphics[width=0.84\linewidth]{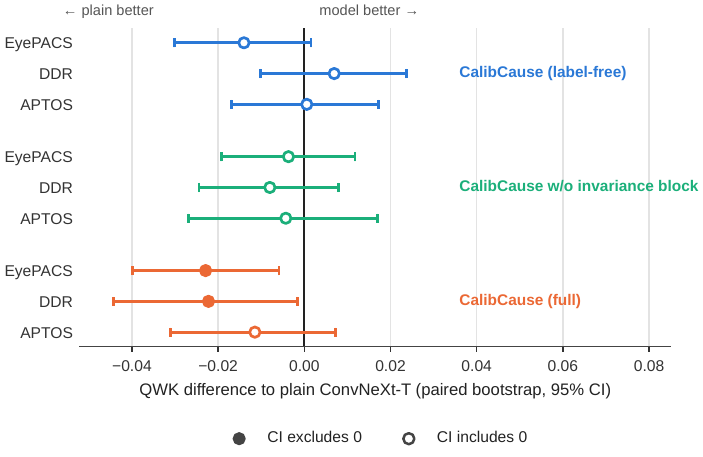}
\caption{Paired QWK difference between each CalibCause variant and the plain ConvNeXt-Tiny on the held-out environment (95\% paired bootstrap interval). Filled markers: interval excludes zero.}\label{fig:paired}
\end{figure}

Table~\ref{tab:loeo} and Fig.~\ref{fig:loeo} compare the four models when each environment is held out in turn. The plain ConvNeXt-Tiny was a strong baseline. Its mean QWK over the three folds (0.743) was the highest, with the label-free model (0.741) and the model without the invariance block (0.738) close behind and the full model lowest (0.724). The paired differences in Fig.~\ref{fig:paired} show that the label-free model could not be told apart from the plain baseline on DDR ($+0.007$, interval $-0.010$ to $+0.024$) or APTOS ($+0.001$), and was slightly lower on EyePACS ($-0.014$, interval $-0.030$ to $+0.002$). The full model was lower than the baseline on EyePACS ($-0.023$) and DDR ($-0.022$), with intervals that exclude zero.

\begin{table}[!tbp]
\caption{Leave-one-environment-out QWK (95\% bootstrap interval) on the held-out environment, seed 42. The last column (s43) gives the held-out EyePACS result for a second seed (43).}\label{tab:loeo}
\footnotesize
\setlength{\tabcolsep}{2.4pt}
\begin{tabular}{@{}lccccc@{}}
\toprule
Model & EyePACS & DDR & APTOS & Mean & EyePACS s43 \\
\midrule
Plain ConvNeXt-T & 0.547 (0.525--0.569) & 0.809 (0.786--0.829) & 0.874 (0.848--0.896) & 0.743 & 0.564 \\
No invariance block & 0.543 (0.522--0.565) & 0.801 (0.777--0.822) & 0.870 (0.843--0.892) & 0.738 & -- \\
Label-free & 0.533 (0.512--0.555) & 0.816 (0.794--0.835) & 0.875 (0.848--0.897) & 0.741 & 0.527 \\
Full & 0.524 (0.500--0.548) & 0.787 (0.762--0.810) & 0.863 (0.834--0.888) & 0.724 & 0.521 \\
\bottomrule
\end{tabular}
\end{table}

The second seed tells how much of this is noise. On held-out EyePACS the plain model moved from 0.547 to 0.564 between seeds, the label-free model from 0.533 to 0.527, and the full model from 0.524 to 0.521. The change between seeds for the plain model (0.017) was larger than its gap to the label-free model in the first seed (0.014), so rankings of single runs on this fold should be read with caution. The ordering of plain above full held for both seeds (paired difference $+0.023$ and $+0.043$, intervals excluding zero). Removing the invariance block (Module~1) did not hurt held-out accuracy: the model without it was at least as accurate as the full model on every fold, and the difference was clear on EyePACS ($+0.019$, interval $+0.004$ to $+0.033$). The invariance objective therefore did not improve cross-site accuracy in our experiments.

\subsection*{What the drop on a new site consists of}
\begin{figure}[!tbp]
\centering
\includegraphics[width=0.9\linewidth]{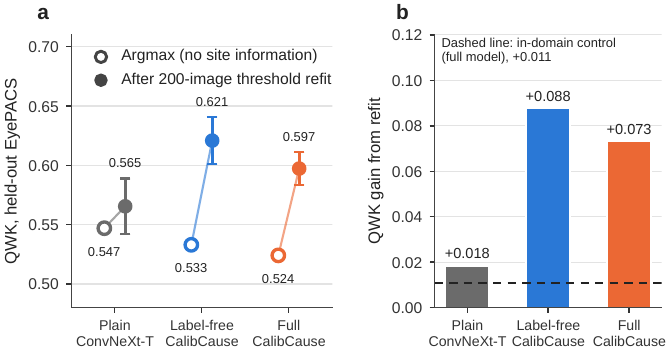}
\caption{Threshold-refit diagnostic on held-out EyePACS. (a) QWK with argmax grading (open) and after refitting four grade thresholds on 200 labelled images from the new site (filled; mean and standard deviation over 30 draws). (b) QWK gain from refitting. The dashed line is the gain for the full model on in-domain data, where no site shift is present.}\label{fig:refit}
\end{figure}

Held-out EyePACS was by far the hardest fold: QWK fell from 0.746 for the in-domain model on the EyePACS test images to 0.524 for the full model trained without EyePACS. This drop mixes at least three things: the loss of about two thirds of the training images, a shift in appearance, and a shift in how EyePACS graders place the boundaries between grades. The threshold-refit diagnostic isolates the last of these (Fig.~\ref{fig:refit}). Refitting four cut-points on 200 labelled images raised held-out QWK from 0.524 to $0.597\pm0.014$ for the full model, from 0.533 to $0.621\pm0.020$ for the label-free model, and from 0.547 to $0.565\pm0.023$ for the plain baseline. The in-domain control for the full model gained only 0.011 (from 0.746 to $0.757\pm0.012$), so the gain on the new site is not an effect of the decision rule alone. After refitting, the label-free model had the highest QWK of the three, reversing the order seen with argmax. We could not run the in-domain control for the plain baseline because no pooled plain model was trained, and a model trained on a size-matched subset was not run, so we cannot say how the remaining gap of about 0.15 to 0.2 divides between lost training data and appearance shift. What the diagnostic does show is that grade boundaries move between sites and that 200 labelled local images recover roughly 0.07--0.09 QWK for CalibCause models.

\subsection*{Expert routing and visual explanations}
\begin{figure}[!tbp]
\centering
\includegraphics[width=0.8\linewidth]{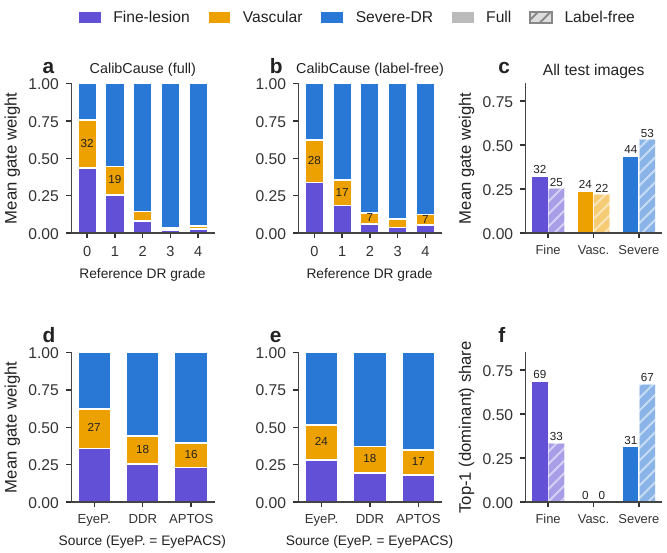}
\caption{Expert routing on the pooled test split. (a, b) Mean gate weight of each expert by reference grade for the full and label-free models. (d, e) The same by source. (c) Mean gate weight over all test images. (f) Share of images for which each expert had the largest weight.}\label{fig:routing}
\end{figure}

The three experts were not used equally (Fig.~\ref{fig:routing}). In the full model the fine-lesion expert was the top-ranked expert for 69\% of test images and the severe-disease expert for 31\%; in the label-free model the shares were 33\% and 67\%. The vascular expert was never top-ranked in either model. Its gate weight was nevertheless not zero: it averaged 0.24 (full) and 0.22 (label-free) and never exceeded 0.35 for any image, so it contributed as a minor, almost constant part of the fused output rather than as a specialist. The gate weight of the severe-disease expert rose with reference grade, from 0.24 (full) and 0.38 (label-free) for grade~0 to 0.95 and 0.88 for grade~4. Routing depended on image quality to a small degree in the full model (correlation between quality score and the weight of the severe-disease expert $-0.17$) and hardly at all in the label-free model ($-0.02$). We therefore describe the gate as severity-driven; we found no evidence that it routes by image quality.

\begin{figure}[!tbp]
\centering
\includegraphics[width=0.68\linewidth]{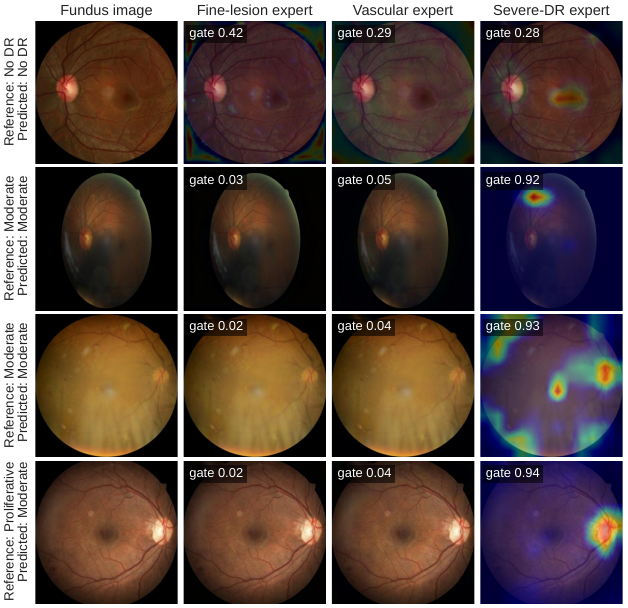}
\caption{Per-expert Grad-CAM maps for the label-free model on four test images, scaled by the gate weight of each expert (shown on each tile). The last row is an error case.}\label{fig:cam}
\end{figure}

\begin{figure}[!tbp]
\centering
\includegraphics[width=0.8\linewidth]{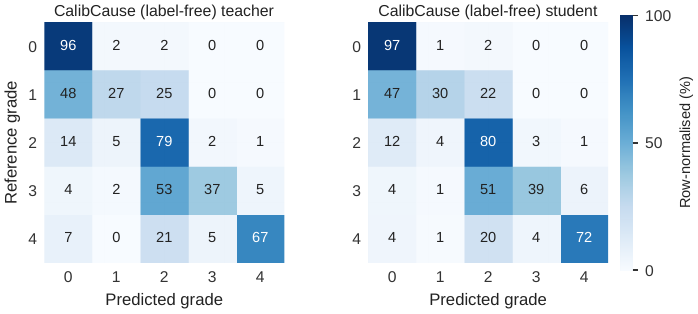}
\caption{Row-normalised confusion matrices for the label-free teacher and student on the pooled test split (percent of each reference grade).}\label{fig:conf}
\end{figure}

Figure~\ref{fig:cam} shows Grad-CAM maps for the label-free model, scaled by gate weight so that an expert that the gate ignores appears faint. For the severe-disease expert the maps concentrate on discrete regions in several images and on the optic disc in another, and for images dominated by that expert the other two maps are close to empty. The maps are qualitative. We did not compare them with lesion annotations, and in the first row the fine-lesion map highlights the image border as well as the retina, so they should be read as indications of where the network looked, not as lesion detections. Figure~\ref{fig:conf} shows that most errors lie between neighbouring grades, with mild disease often graded as normal and severe disease as moderate.

\subsection*{Strengthening invariance and rebalancing routing}
Two further variants of the full model were trained once each (seed 42, 30 epochs). In the first, the adversary was strengthened (environment-loss weight 0.6, gradient-reversal scale 2.0) and a dataset-wide load-balancing loss (weight 0.2) was added. Held-out EyePACS QWK was 0.520, no better than the 0.524 of the full model, and the gate gave the same output for every image (mean weights 0.335, 0.334, 0.331), so the experts were no longer routed by input. In the second, only the load-balancing loss was added. The pooled model then used all three experts (top-ranked for 8\%, 37\% and 55\% of images; pooled QWK 0.856), but held-out EyePACS QWK fell to 0.488. Neither variant is a better candidate than the models above; the vascular expert can be made to participate, but in our runs not at no cost to cross-site accuracy.

\subsection*{Calibration and deferral}
\begin{figure}[!tbp]
\centering
\includegraphics[width=\linewidth]{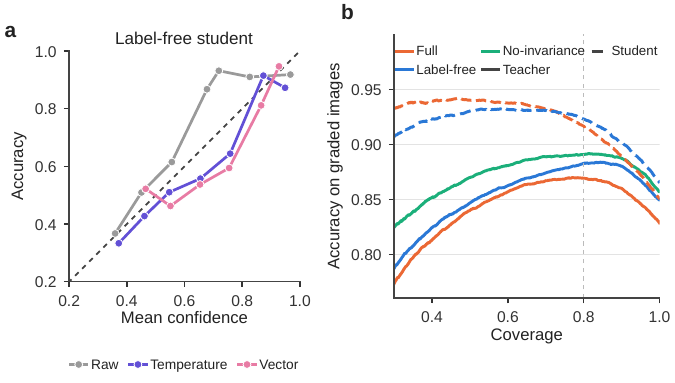}
\caption{Calibration and deferral on the pooled test split. (a) Reliability curve of the label-free student without calibration, after temperature scaling and after vector scaling (ECE of all five models in Table~\ref{tab:pooled}). (b) Accuracy on the images that are graded when the least confident images are deferred to a human reader (coverage $<1$), using temperature-scaled confidence; the dashed vertical line marks 80\% coverage.}\label{fig:cal}
\end{figure}

Without calibration all models were overconfident (ECE 0.17--0.27; Table~\ref{tab:pooled}). Temperature scaling reduced ECE to 0.117--0.149 for the teachers and 0.055--0.057 for the students, and vector scaling to 0.025--0.036 for all five models (Fig.~\ref{fig:cal}a). Under temperature scaling the students were the best calibrated, but under vector scaling the label-free teacher had the lowest ECE (0.025) and the students did not (0.033 and 0.036). The ranking therefore depends on the calibration method, and the differences after vector scaling are about 0.01, so we do not rank the models by calibration. Calibration is also not specific to CalibCause: the same post-hoc step applies to any classifier, and before calibration the plain baseline had similar ECE to the CalibCause models on the held-out EyePACS fold (0.203, against 0.199 for the label-free and 0.220 for the full model).

Confidence ranked images well enough to support deferral (Fig.~\ref{fig:cal}b). Keeping the 80\% most confident test images raised the accuracy of the label-free student from 0.865 to 0.923, that of the full student from 0.851 to 0.916, and that of the label-free teacher from 0.849 to 0.883. Students were the better selective predictors at every coverage between 0.5 and 0.8.

\subsection*{Secondary metrics}
Table~\ref{tab:extra} gives secondary metrics. On the held-out environments the label-free model had a higher referable-DR AUC than the plain baseline on EyePACS ($+0.028$, interval $+0.016$ to $+0.040$) and a higher sensitivity at 90\% specificity ($+0.036$, $+0.014$ to $+0.057$). The model without the invariance block showed a smaller gain on the same fold (AUC $+0.016$; sensitivity $+0.027$), so the gain cannot be attributed to the invariance block. On DDR the intervals included zero, on APTOS the label-free AUC was slightly lower ($-0.009$), and the full model did not differ from the baseline. Averaged over the folds, AUC was 0.919 for the label-free model and 0.910 for the plain baseline. Grading error did not favour the CalibCause models: MAE on held-out EyePACS was higher than for the baseline (label-free $+0.018$, full $+0.032$, intervals excluding zero), and uncalibrated NLL was higher for the label-free model on EyePACS and APTOS and lower on DDR. Macro-F1 is low for all models (0.50 on average on the held-out environments) because it weights every grade equally and grades 1 and 3 were seldom recovered (recall 27--30\% and 37--39\% for the label-free teacher and student, Fig.~\ref{fig:conf}). Better ranking of referable cases without better exact grading is consistent with the shift in grade boundaries described above, since AUC does not depend on the cut-points.

On the pooled test split the label-free student was best on sensitivity at 90\% specificity (0.934), MAE (0.193) and NLL (0.600), tied for best on within-one accuracy (0.949) and below the no-invariance teacher on macro-F1 (0.671 against 0.681). No plain baseline was trained on the pooled split.

\begin{table}[!tbp]
\caption{Secondary metrics from uncalibrated predictions. AUC and sensitivity (Sens@90) are for referable DR; F1 is macro-averaged; MAE is the mean absolute grade error; Within-1 is the fraction of images graded within one grade of the reference; NLL is the negative log-likelihood.}\label{tab:extra}
\small
\setlength{\tabcolsep}{4pt}
\begin{tabular}{@{}lcccccc@{}}
\toprule
Model & AUC & Sens@90 & F1 & MAE & Within-1 & NLL \\
\midrule
\multicolumn{7}{@{}l}{\emph{Held-out environment, mean over three folds}} \\
Plain ConvNeXt-T & 0.910 & 0.835 & 0.503 & 0.384 & 0.902 & 1.240 \\
No invariance block & 0.914 & 0.854 & 0.511 & 0.395 & 0.901 & 1.279 \\
Label-free & 0.919 & 0.853 & 0.511 & 0.380 & 0.898 & 1.265 \\
Full & 0.908 & 0.839 & 0.495 & 0.410 & 0.897 & 1.279 \\
\midrule
\multicolumn{7}{@{}l}{\emph{Pooled test split}} \\
Full, teacher & 0.938 & 0.881 & 0.605 & 0.255 & 0.930 & 0.846 \\
Label-free, teacher & 0.945 & 0.905 & 0.648 & 0.218 & 0.941 & 0.780 \\
No invariance block, teacher & 0.944 & 0.917 & 0.681 & 0.201 & 0.949 & 0.719 \\
Full, student & 0.968 & 0.922 & 0.638 & 0.217 & 0.942 & 0.630 \\
Label-free, student & 0.969 & 0.934 & 0.671 & 0.193 & 0.949 & 0.600 \\
\bottomrule
\end{tabular}
\end{table}

\subsection*{Per-source training and external testing}
Trained and tested within one source, as in QGDR, the full model reached QWK 0.9345 on DDR (accuracy 0.909; QGDR as reported: QWK 0.8231), 0.9137 on APTOS and 0.7607 on EyePACS (Table~\ref{tab:persource}). The EyePACS figure is not comparable with QGDR, which used a smaller EyePACS subset with quality labels. On DDR, the only dataset with a within-source result in both studies, Table~\ref{tab:qgdr} compares the metrics that QGDR reports: our values are higher on every accuracy-type metric, and for ECE the order depends on the calibration step. The comparison is indicative only, since QGDR used $224\times224$ images and its own random split, we used $384\times384$ images and the split described in Methods, and we did not train a plain network in the per-source setting, so the gap cannot be assigned to the quality-gated design. On the external sets the full teacher reached QWK 0.764 on IDRiD and 0.825 on DeepDRiD, with sensitivity for referable DR of 0.926 and 0.978; accuracy was lower (0.624 and 0.688) and ECE was 0.130 and 0.195.

\begin{table}[!tbp]
\caption{Per-source and external results for the full CalibCause model (teacher). Referable DR is grade $\ge2$.}\label{tab:persource}
\small
\setlength{\tabcolsep}{4pt}
\begin{tabular}{@{}llccccc@{}}
\toprule
Setting & Test set & Acc & QWK & AUC & Sens & Spec \\
\midrule
Per-source & DDR & 0.909 & 0.935 & 0.964 & 0.934 & 0.957 \\
Per-source & APTOS & 0.862 & 0.914 & 0.962 & 0.963 & 0.940 \\
Per-source & EyePACS & 0.829 & 0.761 & 0.913 & 0.741 & 0.967 \\
External & IDRiD & 0.624 & 0.764 & 0.954 & 0.926 & 0.839 \\
External & DeepDRiD & 0.688 & 0.825 & 0.973 & 0.978 & 0.818 \\
\bottomrule
\end{tabular}
\end{table}

\begin{table}[!tbp]
\caption{Within-source results on DDR: QGDR as reported in its paper~\cite{pan2026qgdr} and the full CalibCause model (this study). Referable DR is grade $\ge2$. Macro AUC is the one-vs-rest AUC over five grades (our computation; QGDR's definition may differ). CalibCause ECE is shown raw and after temperature (T) and vector (V) scaling. Splits and image sizes differ between the studies.}\label{tab:qgdr}
\small
\setlength{\tabcolsep}{5pt}
\begin{tabular}{@{}lcc@{}}
\toprule
Metric & QGDR & CalibCause (full) \\
\midrule
Accuracy & 0.809 & 0.909 \\
QWK & 0.823 & 0.935 \\
Macro AUC (5-class) & 0.902 & 0.943 \\
Referable AUC & 0.952 & 0.964 \\
Referable sensitivity & 0.882 & 0.934 \\
Referable specificity & 0.877 & 0.957 \\
Referable F1 & 0.868 & 0.941 \\
Referable precision & 0.854 & 0.947 \\
ECE & 0.066 & 0.186 (raw), 0.117 (T), 0.020 (V) \\
\bottomrule
\end{tabular}
\end{table}

\section{Discussion}\label{sec:discussion}

We set out to combine properties that DR screening needs but that are rarely studied together: awareness of image quality, inspectable routing and visual evidence, no dependence on manual quality labels, a compact deployable model and usable confidence. These can be had without losing accuracy against a plain ConvNeXt-Tiny trained with the same recipe, but in our experiments they did not improve cross-site accuracy either: averaged over three held-out environments the plain baseline scored 0.743 QWK and the label-free model 0.741.

The adversarial invariance block is the clearest negative result. The model without it matched or exceeded the full model on every held-out fold, and the full model was below the plain baseline on two folds. Adversarial invariance across three sources, one much larger than the others, may be too weak a signal to remove acquisition cues yet strong enough to cost useful detail; strengthening the adversary did not help. We therefore no longer describe the framework as site-invariant. The label-free variant, which keeps the block, did better than the full model, but without a label-free model lacking the block we cannot attribute that difference to either change.

The drop on held-out EyePACS is informative beyond the model comparison. Part of it is a shift in grade boundaries: refitting four thresholds on 200 labelled images recovered 0.07--0.09 QWK for the CalibCause models, and an in-domain control showed this was not an artefact of the decision rule. This agrees with reports that grader variability and the reference standard affect what a model learns~\cite{krause2018}. The remainder, which we cannot divide between lost training data and appearance shift, is still large. A programme adopting a model trained elsewhere should expect to recalibrate its grade boundaries locally from a labelled sample of a few hundred images. The smaller gain of the plain baseline from refitting (0.018 against 0.073 and 0.088) rests on one fold and one seed, and we draw no conclusion from it. The secondary metrics point the same way: the label-free model ranked referable cases better than the baseline on held-out EyePACS but did not grade more exactly.

The student is the most practically useful result. Distillation from the label-free teacher gave a network the size of the plain baseline that was more accurate than its teacher and a better selective predictor. We did not train the plain baseline on the pooled split, so we cannot say whether distillation from the quality-gated teacher beats training the same architecture directly. Student training was fragile: with one learning rate for all layers it collapsed to the majority class, and it trained reliably only after the pretrained backbone received a tenth of the head learning rate, with warm-up and gradient clipping.

The routing analysis corrects an impression that the architecture alone might give. The vascular expert never had the highest weight, and the gate depended mostly on severity and little on image quality, so the output is better described as a two-expert system with a small, nearly constant vascular contribution. A load-balancing loss made the vascular expert active in a pooled model but at a cost on the held-out site, and we did not test removing the expert altogether. Gate-weighted Grad-CAM maps make the routing visible, but remain qualitative.

On calibration our claim is limited. After vector scaling the five models were within about 0.01 ECE of each other, and calibration applies equally to a plain classifier. The useful result is that the calibrated confidence of the students ranked images well enough that deferring a fifth of them raised accuracy on the rest by about 6 percentage points; as with the grade boundaries, the deferral threshold must be set locally.

\paragraph{Limitations.} All data are public research collections and not a consecutive screening population, so prospective testing is still needed. The two additional variants were run once, and a second seed was run only for held-out EyePACS and for three of the four models; the bootstrap intervals reflect test-set sampling only. There was no pooled plain baseline, no size-matched control for the held-out-site drop and no in-domain refit control for the plain model, and external testing was run for the full model only. Vector-scaled ECE, confusion matrices and secondary metrics were computed from saved predictions with one script applied to all models. The label-free and full students were trained with different recipes, and Grad-CAM maps were not validated against lesion annotations. The threshold refit needs 200 labelled images from the target site and is not zero-shot. Gradient reversal, distillation, Grad-CAM and post-hoc calibration are established methods; our contribution is a quality-gated pipeline that combines them and a test of what the combination does and does not add over a plain network.

Related work on graph-based DR grading~\cite{alamri2026ensemble}, quality- and uncertainty-aware DR classifiers~\cite{akter2025graphuncertainty,rejectionunc2025}, domain generalisation for DR grading~\cite{zhangliu2025phasedg}, interpretable mixture-of-experts routing~\cite{cheng2025ermoe} and mixture-of-experts models for medical images~\cite{chen2024lowrank} addresses parts of this problem. Shortcut learning, in which a network relies on cues that do not reflect the disease~\cite{geirhos2020,degrave2021}, is the failure that the invariance block was meant to reduce, and our results suggest that an adversarial block alone is not enough to do so across three sources.

Future work should include a pooled plain baseline, repeated seeds on every fold, a tuned or alternative invariance objective, routing variants that give the vascular expert a role, and prospective evaluation on consecutive screening images with local recalibration of the grade boundaries.

\section*{Acknowledgements}
The authors thank the Department of Data Science and Artificial Intelligence, National University of Computer and Emerging Sciences (FAST-NUCES), Islamabad, for institutional support.

\section*{Conflict of Interest}
The authors declare no conflict of interest.

\section*{Funding}
This research received no specific grant from any funding agency.

\section*{Data Availability}
This study used publicly available datasets: EyePACS, DDR, APTOS~2019, IDRiD and DeepDRiD. All are accessible through the repositories cited in the references.

\section*{Ethics Approval}
This study used only publicly available, de-identified retrospective image datasets and did not involve new data collection from human participants; separate ethics approval was therefore not required.

\setlength{\bibsep}{1pt plus 0.3ex}
\bibliographystyle{unsrtnat}
\bibliography{sn-bibliography}

\end{document}